\documentclass[twoside,12pt]{article}
\usepackage{epsfig}

\newcommand{\be}{\begin{equation}}
\newcommand{\ee}{\end{equation}}
\newcommand{\bea}{\begin{eqnarray}}
\newcommand{\eea}{\end{eqnarray}}

\begin{document}

\title{ \vspace{1cm} Hadrons in Nuclei\\ from High (200 GeV) to Low (1 GeV) energies}
\author{K.\ Gallmeister, M.\ Kaskulov, U.\ Mosel\footnote{mosel@physik.uni-giessen.de},
 P.\ M\"uhlich\\
Institut fuer Theoretische Physik, Universitaet Giessen, Giessen,
Germany}

\maketitle
\begin{abstract}
The study of the interaction of hadrons, produced by elementary probes in
a nucleus, with the surrounding nuclear medium can give insight into two
important questions. First, at high energies, the production process, the
time-scales connected with it and the prehadronic interactions can be
studied by using the nuclear radius as a length-scale. We do this here by
analyzing data from the EMC and HERMES experiements on nuclear
attenuation. Second, at low energies the spectral function, and thus the
selfenergy of the produced hadron, can be studied. Specifically, we
analyze the CBELSA/TAPS data on $\omega$ production in nuclei and discuss
the importance of understanding in-medium effects both on the primary
production cross section and the final state branching ratio. In both of
these studies an excellent control of the final state interactions is
essential.
\end{abstract}

\section{Introduction}
The investigation of hadrons in medium is an exciting subject with
implications both for hadron physics and the study of the quark-gluon
plasma (QGP) state of nuclear matter. The former field of studies is
concerned with the interactions of slow-moving hadrons and their
selfenergies inside nuclear matter. The latter deals with interactions of
fast-moving partons and/or hadrons inside hot and dense nuclear matter;
their clarification needs both the properties of the still largely unknown
QGP state of nuclear matter and of the interaction of jets with this
matter. A study of jet interactions in cold, ordinary nuclear matter of
known properties may thus help to disentangle effects of the interaction
of jets from those of the medium in which they move. Also, at these high
energies the question of production and formation times may be studied by
exploiting the final state interactions of the hadrons produced with the
surrounding nuclear matter. For this problem the nuclear radius provides a
length- and thus also a time-scale during which interactions can take
place.

In this paper we briefly describe our studies of the formation times
and the interactions during this time. For this purpose we first
analyze data both from the EMC ($\approx 200$ GeV) and from the
HERMES collaboration ($\approx 20$ GeV). In the second part we then
move to a discussion at much lower energies (a few 100 MeV) and
discuss our present knowledge of in-medium selfenergies of hadrons
and their interactions. For details for the former we refer to
\cite{GM07} whereas for the latter we mainly refer to
\cite{Muehlich07}.

\section{Prehadronic Interactions}
The EMC experiment has yielded data on nuclear attenuation that
correspond to average values of $Q^2 \approx 4 - 12$ GeV$^2$,
corresponding to the two beam energies of 100 GeV and 280 GeV,
respectively. However, most of the experiments on nuclear
attenuation that have been performed (HERMES, JLAB) work at rather
small values of $Q^2 \approx 1 - 2$ GeV$^2$ so that methods of
perturbative QCD are not applicable. We have, therefore, modelled
the prehadronic interactions such that the description works at all
energy regimes and describes the transition from high to low
energies correctly. Our model relies on a factorization of hadron
production into the primary interaction process of the lepton with a
nucleon, essentially taken to be the free one, followed by an
interaction of the produced hadrons with nucleons.

For the first step we use the PYTHIA model that has been proven to
be very successful for describing hadron production, also at the low
values of $Q^2$ and $\nu$ treated in this paper. This model contains
not only string fragmentation, but also direct interaction processes
such as diffraction and vector meson dominance. In this first step
we take nuclear effects such as Fermi motion, Pauli blocking and
nuclear shadowing into account \cite{Falter}. The relevant
production and formation times are obtained directly from PYTHIA
\cite{KGTF}; for a definition of these times we refer to
\cite{GM07}. In the second step we introduce prehadronic
interactions between the production and the formation time and the
full hadronic interactions after the hadron has been formed. We do
this by means of the semiclassical GiBUU transport code \cite{GiBUU}
which not only allows for absorption of the newly formed hadrons,
but also for elastic and inelastic scattering as well as for
side-feeding through coupled channel effects.

The actual time-dependence of the prehadronic interactions presents
an interesting problem in QCD. Dokshitzer \cite{Dok} has pointed out
that QCD and quantum mechanics lead to a time-dependence somewhere
between linear and quadratic. Kopeliovich et al. \cite{Kopel} have
called attention to the fact that the cross sections may even
oscillate in time, as a consequence of a quantum mechanical
superposition of excited states of the produced hadron. We also note
that a linear behavior had been used by Farrar et al.\ \cite{Farr}
in their study of quasiexclusive processes. In our calculations we
work with different time-dependence scenarios, among them a
constant, lowered prehadronic cross section, a linearly rising one
and a quadratically rising one. In addition, we study a variant of
the latter two, where the cross section for leading hadrons, i.e.
hadrons that contain quarks of the original target nucleon, starts
from a pedestal value $\sim 1/Q^2$, thus taking into account
possible effects of color transparency (for details see
\cite{GM07}).

\begin{figure}[htb]
\begin{center}
\epsfig{file=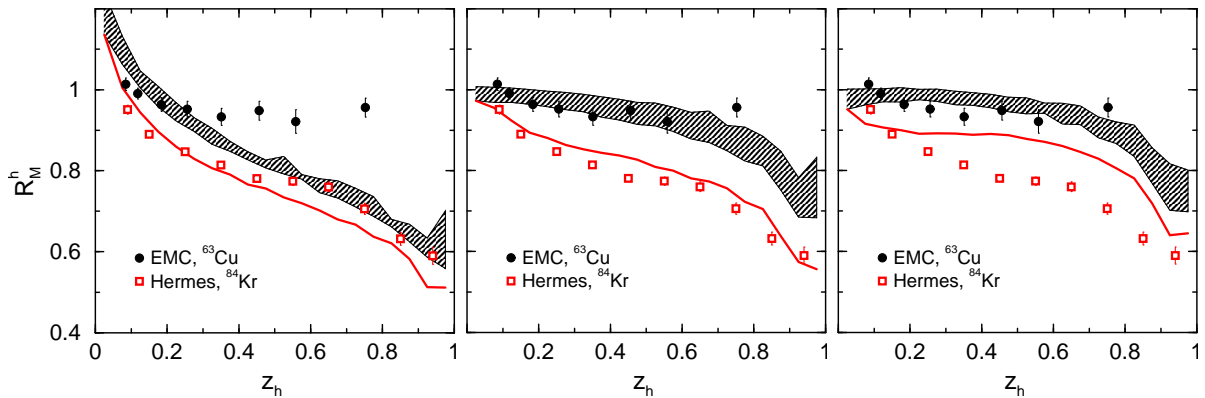} \caption{Nuclear modification factor for
charged hadrons. Experimental data are shown for HERMES at 27 GeV
and for EMC at 100/280 GeV. For the latter the predictions for the
upper and lower energy are given by the upper and lower bounds of
the shaded band. The cross section scenarios are (from left to
right): constant, linear and quadratic increase with time after
production.} \label{att}
\end{center}
\end{figure}
Fig. \ref{att} shows a comparison of these various model
assumptions. It is clearly seen that the assumption of prehadronic
cross section reduced to a constant value (=0.5) in the leftmost
figure leads to a significantly too large attenuation for the EMC
experiment while the HERMES data can be described reasonably well
(this should be no surprise since the constant has essentially
been fitted to the HERMES data). On the contrary, the quadratic
time-dependence gives a good description of the EMC data but
underestimates the attenuation significantly for the HERMES data.

\begin{figure}[htp]
\begin{center}
\epsfig{file=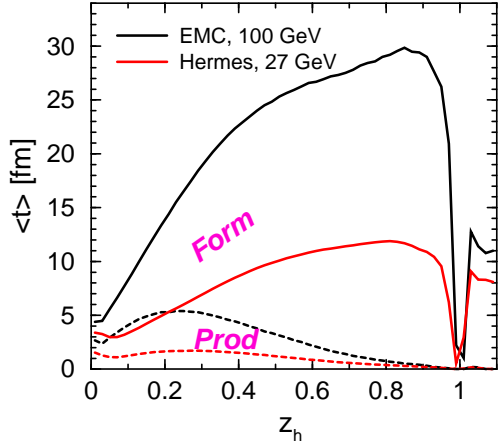, width=6cm, angle=-90} \caption{Averaged
production (lower 2 curves) and formation times (upper 2 curves)
for the EMC (solid) and the HERMES (dashed) experiment. The
average contains the times of the leading hadrons as a function of
the variable $z_h$, defined as the ratio of the hadron's energy to
the energy transfer.} \label{times}
\end{center}
\end{figure}

\begin{figure}[htb]
\begin{center}
\epsfig{file=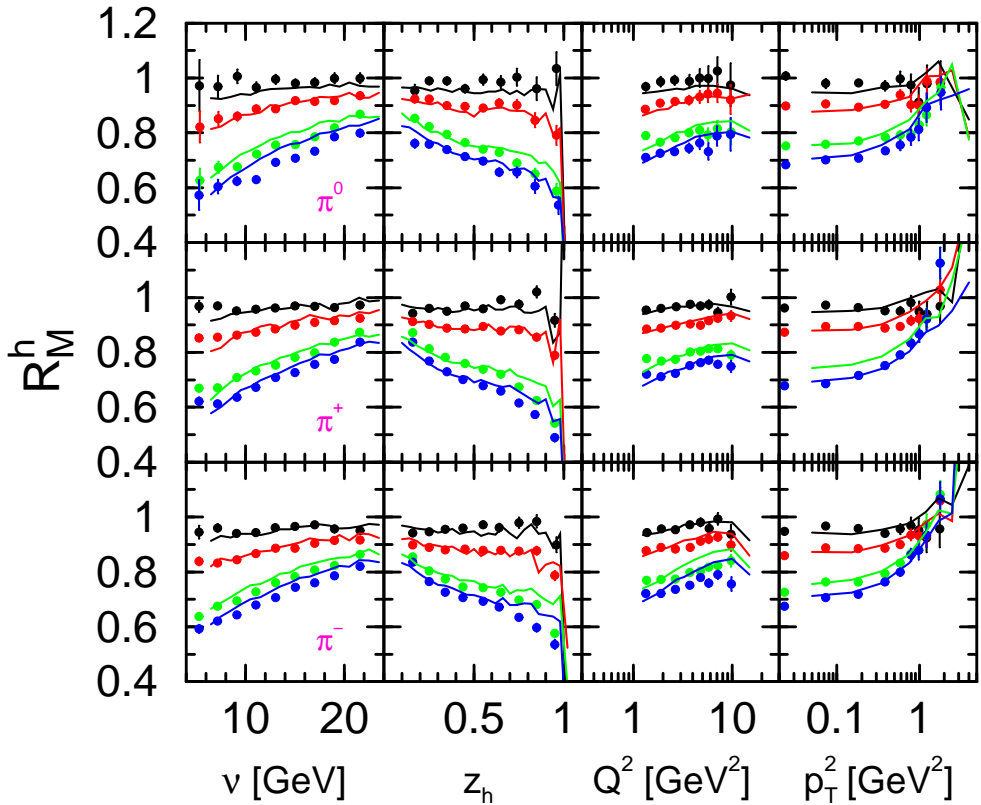, width=13cm} \caption{Attenuation of
pions in comparison with results of the HERMES experiment. Target
nuclei are $^4He, ^{20}Ne, ^{84}Kr$ and $^{131}Xe$ (curves from top
to bottom). The calculations have been done with the linear increase
of the prehadronic cross sections and a pedestal value for the
leading hadrons $\sim 1/Q^2$. Experimental acceptance limitations
are taken into account. Data are from \cite{Hermes}.}
\label{pionatt}
\end{center}
\end{figure}

This behavior can be understood by looking at the relevant
production and formation times depicted in Fig.\ \ref{times}. One
sees that the formation times for the EMC experiment at $z_h
\approx 0.8$ approach 30 fm, i.e. values well beyond nuclear
radii. This then causes very little attenuation if the prehadronic
cross section does not rise fast enough so that the hadron still
interacts with the nuclear environment.
The middle scenario in Fig.\ \ref{att} (linear time-dependence)
describes the data for all hadrons both for the high-energy EMC
experiment and for the lower-energy HERMES experiment very well.
This nearly perfect agreement is also seen in Fig.\ \ref{pionatt}
which gives the attenuation $R$ for pions for the same target nuclei
as in Fig.\ \ref{att} as a function of energy-transfer $\nu$,
relative energy $z_h = E_h/\nu$, momentum transfer $Q^2$ and the
squared transverse momentum $p_T^2$. The dependence of $R$ on all
these dynamical variables is described very well. The rise of $R$
with $\nu$ is mainly an acceptance effect, as we have shown in
\cite{Falter}, whereas the weaker rise of $R$ with $Q^2$ reflects
the pedestal value $\sim 1/Q^2$ of the prehadronic cross sections.
The $z_h$-dependence of $R$ is -- below $z_h \approx 0.5$ -- strongly
influenced by final state interactions of fast-moving hadrons that get
slowed down by collisions.

\begin{figure}[htb]
\begin{center}
\epsfig{file=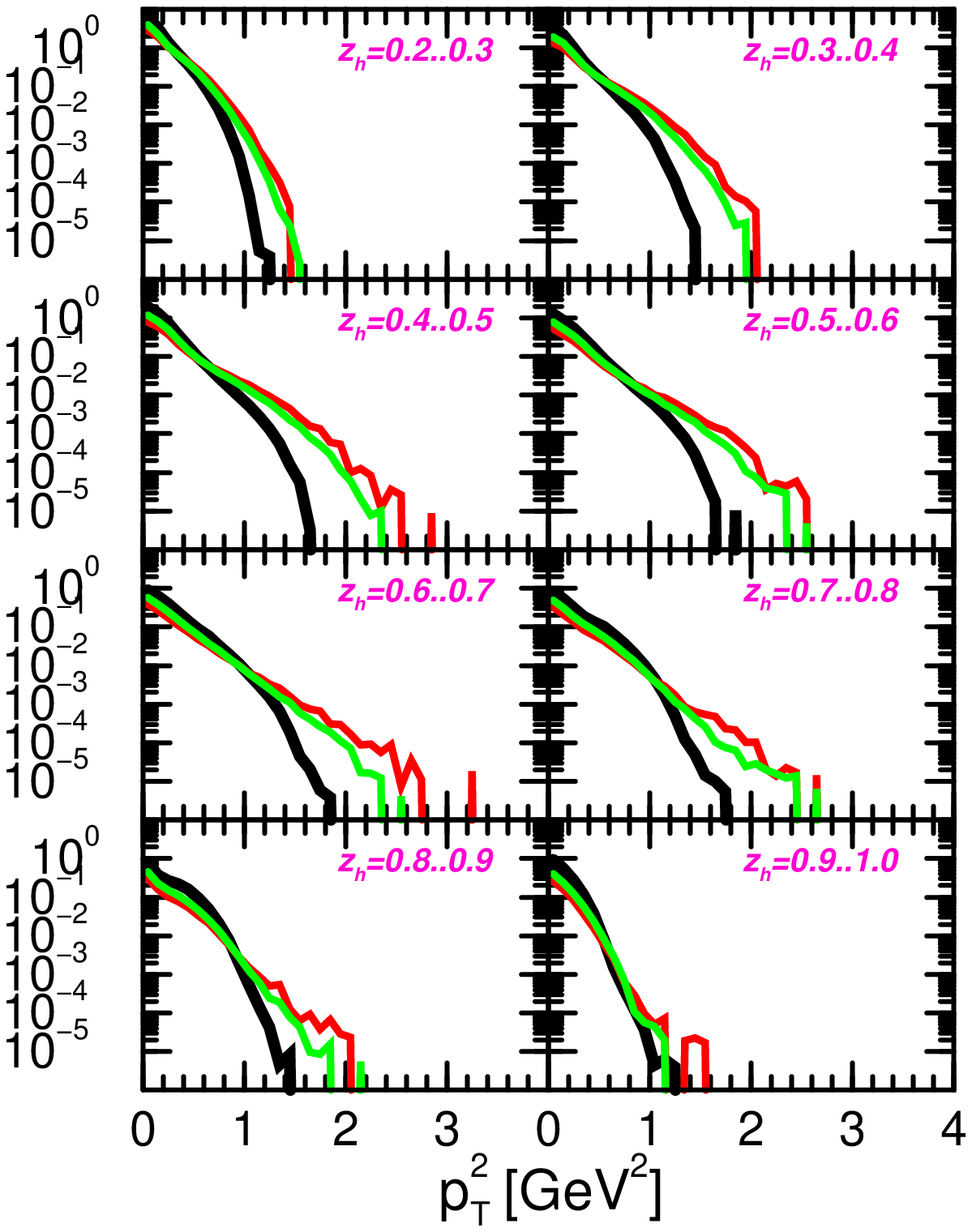, width=8.5cm} \caption{Transverse momentum
spectra (arbitrary units) of $\pi^+$  production on a $Pb$ target for an
electron energy of 5 GeV within the CLAS acceptance for different $z_h$
bins. The solid line gives the distribution for a $D$ target for
comparison; the middle curve shows the results without and the top curve
those with prehadronic interactions.} \label{pt}
\end{center}
\end{figure}

The curves in the rightmost column of Fig.\ \ref{pionatt} show that for
the heavier target nuclei the transverse momentum distribution of the
nuclear modification factor $R$ tends towards values $>1$ at large
$p_T^2$. We investigate this phenomenon in more detail with the help of
the results obtained for the lower JLAB energy of 5 GeV, shown in Fig.\
\ref{pt}.
Here it is seen that the $p_T$ distributions reach farther out for heavier
nuclei than for a light $D$ target. This is a consequence of hadronic
rescattering and obviously leads to the observation that $R > 1$ at these
high $p_T$. That the $p_T$ distributions reach out farthest for
intermediate $z_h$ values is due to simple kinematic constraints.
Prehadronic interactions have little effect on this observable as can be
seen by comparing the upper two curves (with and without prehadronic
interactions) with each other. It can also be seen that the results for
$Pb$ and for $D$ agree with each other for low to moderate transverse
momenta. Only at the highest transverse momenta a nuclear effect appears.
As a consequence, the expectation values $\langle p_T^2 \rangle$ are
hardly affected by rescattering and are thus not a good indicator for any
rescattering (be it hadronic or prehadronic) effects.

With this method we have also calculated the recently measured
transmission of large $z$ pions moving in forward direction through a
nucleus \cite{Clasie}. Again a very good agreement of theory and
experiment is reached \cite{Kask}.


Summarizing this section we can say that our event simulation gives a very
good description of nuclear attenuation data over a wide (5 GeV - 200 GeV)
energy range. The data are determined by prehadronic and hadronic final
state interactions. A consistent description of all these data can be
obtained only if the prehadronic interaction strength grows linearly with
time.

\section{Hadronic Spectral Functions}

In elementary reactions at lower energies and in ultrarelatistic heavy ion
collisions the focus has always been on reconstructing the spectral
function of hadrons in medium from the four-vector distributions of their
decay products which are ultimately seen in the detector. Little attention
has been paid to the fact that the observed invariant mass distribution is
given by (in its simplest form) a product of production cross section
$\sigma_{\rm in}$, spectral function $\mathcal{A}$ and decay branching
ratio $B_{\rm out}$ which all three depend on the invariant mass $\mu$ of
the produced meson
\be
\frac{d\sigma}{d\mu^2} \approx \sigma_{\rm in} (\mu)\, \mathcal{A}(\mu)
B_{\rm out}(\mu) = \sigma_{\rm in} (\mu)\, \frac{1}{\pi}
\frac{\mu\,\Gamma_{\rm tot}(\mu)}{(\mu^2 - m^2)^2 + \mu^2 \Gamma_{\rm
tot}^2(\mu)} \, B_{\rm out}(\mu)~.
\ee
The production cross section $\sigma_{\rm in}$ contains a phase-space
factor that naturally favors the production of low-mass $\omega$'s. In
addition, there is a sizeable final state interaction if one or both of
the decay products are hadrons.

In ultrarelativistic heavy ion collisions the emission often
happens from a thermalized state so that the mass-dependence of
the production plays no role and the decay branching ratio is
known if the experiment looks for dileptons, as the heavy ion
experiments HADES, CERES and NA60 \cite{Had,CERES,NA60} do. These
reactions have the advantage that high densities and/or
temperatures are reached. They have the inherent disadvantage,
however, that any experimental signal necessarily contains a
time-integral over the whole reaction and thus states with
different characteristics (preequilibrium, equilibrium at varying
temperatures and densities, different states of matter (hadronic
of QGP)). Any analysis of results from such reactions thus needs a
description of the reaction dynamics that is as good as that of
the static equilibrium in-medium properties of the hadron.

In contrast, in reactions with elementary probes on nuclei
\cite{g7,TAPS,KEK} most of the densities probed are below that of
nuclear matter, but the nuclear state remains much more stationary
and the medium is thus much better defined. However, in this case
the primary production cross section may be strongly dependent on
the mass of the produced hadron, in particular if the experiment is
performed with bombarding energies close to threshold. This is, for
example, the case for the CBELSA/TAPS experiment \cite{TAPS,Metag}
that exploits the reaction $\gamma + A \to \pi^0 + \gamma + A^*$ to
look for the in-medium spectral function of the $\omega$ meson. This
experiment uses incoming photons that cover an energy range from
below the free $\omega$ production threshold up to energies well
above. In addition, in this experiment not only do the outgoing
pions experience strong final state interactions, but also the
branching ratio is strongly dependent on the invariant mass. This is
easy to see since just under the $\omega$ peak the decay channel
$\omega \to \rho \pi$ opens up so that the in-medium properties of
the $\rho$ meson do influence those of the $\omega$ meson.

\begin{figure}[htb]
\begin{center}
\epsfig{file=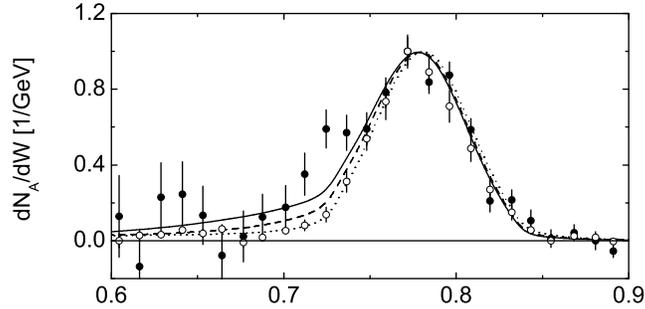, width=9cm} \caption{Comparison of
calculated $\omega$ spectral function as a function of the
invariant mass (in GeV) in comparison to the data from \cite{TAPS}
in the photon energy range $900 - 2200$ MeV (open symbols: data on
$H$ target, solid: data on $Nb$ target). The dotted curves
contains only collisional broadening, the dashed curve contains
the collisional broadening together with a downward mass shift of
8\%, the solid curve finally contains broadening and a 16\%
downward mass shift (from \cite{MDiss}).} \label{subthr}
\end{center}
\end{figure}

We have shown earlier that the GiBUU method describes the experimentally
obtained spectral function very well \cite{MFM} if a lowering of the
$\omega$ mass by 14 - 16\% at saturation density is put in by
hand\footnote{Note that recent calculations of the $\omega$ spectral
function in medium give only a significantly smaller mass shift
\cite{VS,LWF}.}. GiBUU is a semiclassical event generator that has been
developed to describe a wide range of nuclear reactions, from
neutrino-induced over electroproduction reactions to heavy-ion reactions,
all with the same physics input and the same code \cite{GiBUU}. In this
sense it is unique. Since the method has been extensively tested against
data in all these different reactions we feel confident that we can now
use it to explore the physics of the effects observed in the CBELSA/TAPS
experiment. Fig. \ref{subthr} shows the results of such a study.
Immediately noticeable is that the calculated spectral function with all
in-medium corrections included shows a broadening only on the low-mass
side of the peak; the right side is nearly unaffected by the mass-shift
and the broadening.
The explanation for this observation is that the data have been
taken in an energy-interval $900 - 2200$ MeV that covers the free
$\omega$ production threshold at $E_{\rm thresh} = 1100$ MeV. Since
the simulation contains an explicit mass shift in the spectral
function this also shifts, correspondingly, the threshold for
$\omega$ production down. This downward shift of the threshold is
the reason for the observed enhancement at lower masses.

\begin{figure}[htb]
\begin{center}
\epsfig{file=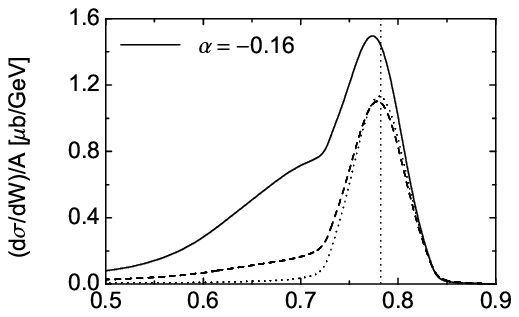, width=7cm}
\epsfig{file=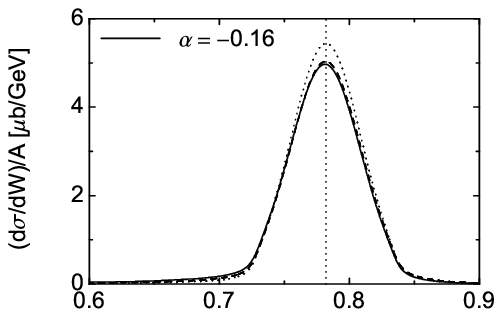, width=7cm} \caption{Invariant mass
spectrum in the $\pi^0 \gamma$ channel from $^{40}Ca$ in the
photon energy range 900 - 1200 MeV (left) and 1500 - 2000 GeV
(right). The results contain a folding with a mass resolution of
25 MeV and a $1/E_\gamma$ weighting of a bremsstrahlung spectrum.
The dotted lines give the vacuum spectral function, the dashed
lines contain collisional broadening and the solid curves an
attractive mass shift of 16\% at saturation density (from
\cite{MDiss}).} \label{thresheffect}
\end{center}
\end{figure}
In order to substantiate this explanation we show in Fig.\
\ref{thresheffect} a comparison of the results of simulations for
the reaction $\gamma + ^{40}Ca \to \pi^0 X$ in the photon energy
intervals $900 - 1200$ MeV (left), which covers the free $\omega$
production threshold, and $1500 - 2000$ GeV (right) which is
clearly above the threshold.

Fig.\ \ref{thresheffect} shows clearly that the low energy
calculation (left) yields a very large surplus of strength on the
low-mass side of the $\omega$ peak, whereas at the higher energy
nearly no enhancement can be seen. At the higher bombarding energy
faster and faster $\omega$ mesons are produced so that more and
more of them will decay outside the nuclear target, with a free
spectral function. Only with a very restrictive cut on $\omega$
momenta, that enriches the in-medium decays, an effect could again
be seen.

\begin{figure}
\begin{center}
\epsfig{file=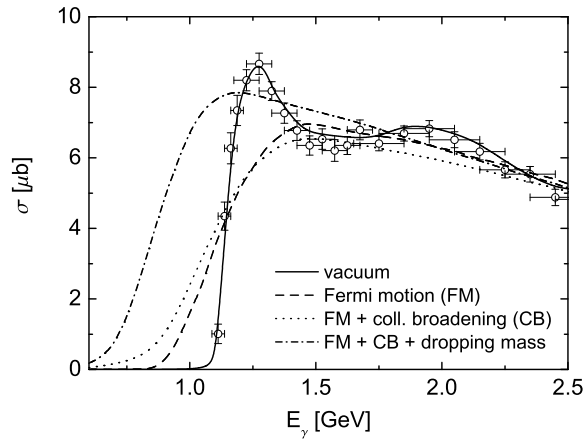, width=8cm} \caption{Total exclusive
$\omega$ production cross section. The curves including medium
modifications are calculated at saturation density $\rho_0$. Data
from \cite{omdat} (from \cite{MDiss}).} \label{omel}
\end{center}
\end{figure}

This strongly suggests that also the experimentally seen
enhancement at the lower masses is due to a threshold effect. The
production of $\omega$ mesons in medium -- with a lowered spectral
function -- is in the calculations described by first extracting
the $s$-dependence of the matrix element from the free
experimental cross section \cite{omdat}
\begin{equation}
\sigma_{\gamma N \to \omega N}^{\rm exp}(s) = \frac{1}{16\pi s
|\mathbf{k}_{\rm cm}|} \int^{(\sqrt{s} - m_{N})^2}_{m_\pi^2}
d\mu^2 |\mathcal{M}_{\gamma N \to \omega N} (s)|^2
\mathcal{A}_\omega (\mu,\rho=0) |\mathbf{q}_{\rm cm}(\mu)|
\end{equation}
under the assumption that the dependence of $\mathcal{M}$ on the
spectral function $\mathcal{A}$ of the $\omega$ meson in vacuum
can be neglected. This gives the elementary cross section for
$\gamma + N \to \omega N$ shown in fig.\ \ref{omel}.
In a second step then, for use in the medium, the $s$-dependence
of the matrix element is scaled (for details see \cite{MDiss}).
After integrating over the Fermi motion and inserting an in-medium
spectral function this gives the dash-dotted curve in Fig.\
\ref{omel}. The difference between the solid (vacuum) and the
dash-dotted (in-medium) curve is responsible for the effect seen
in the CBELSA/TAPS experiment. If this treatment of the in-medium
amplitude is sufficient for a realistic description of threshold
effects will have to be more closely investigated now that the
origin of the observed enhancement has been clarified.

\section{Summary}
Summarizing we emphasize again that any extraction of formation
times and prehadronic interactions from nuclear attenuation data
requires a state-of-the-art treatment of final state interactions
that is as reliable as the QCD input. The same holds for
experimental determinations of hadronic in-medium spectral functions
where -- besides the final state interactions -- also the initial
production process may play a significant role.

We gratefully acknowledge the discussions with and help of the
whole GiBUU team: O. Buss, A. Larionov, T. Leitner, B.
Steinmueller and J. Weil. This work has been supported by BMBF and
DFG.

\end{document}